# Establishing and Measuring Standard Spreadsheet Practices for End-Users

Garry Cleere
garry.cleere@q-validus.com
*Q-Validus Ltd.*
*NovaUCD*
*Belfield Innovation Park*
*University College Dublin*
*Belfield, Dublin 4, Ireland*

## ABSTRACT

This paper offers a brief review of cognitive verbs typically used in the literature to describe standard spreadsheet practices. The verbs identified are then categorised in terms of Bloom's Taxonomy of Hierarchical Levels, and then rated and arranged to distinguish some of their qualities and characteristics. Some measurement items are then evaluated to see how well computerised test question items validate or reinforce training or certification. The paper considers how establishing standard practices in spreadsheet training and certification can help reduce some of the risks associated with spreadsheets, and help promote productivity.

## 1. INTRODUCTION

Organisations and managers need to know that spreadsheet end-users are sufficiently competent and careful in their work with business critical spreadsheets. This paper considers a range of standard practices for spreadsheet end-users drawn from various sources. The paper looks at the kinds of language used, (cognitive verbs in particular), to describe the standard practice approaches, and then considers how the kinds of competencies or behaviours entailed by these verbs may be satisfactorily measured.

A number of documents are reviewed including various recommended practice materials, some EuSpRIG papers, and books, as well as the Syllabus content for the Spreadsheet Safe™ course certification. Sources reviewed in the literature are included in the references.

The Spreadsheet Safe™ programme promotes standard practices in spreadsheets end-use as a way to help reduce risks for organisations and promote productivity. As such Spreadsheet Safe™ is a valuable measure for managers who need to know their teams are skilled in applying standard spreadsheet usage practices, and can work productively. The certification is a flexible, full-solution offering which includes a manual, e-learning, and an online certification test. Spreadsheet Safe™ provides a certificate of competence for Candidates who wish to show that they are skilled, competent and responsible in their work with significant or business critical spreadsheets.

It is now recognized that many organisations are unaware that their corporate spreadsheets or models are not designed, maintained or controlled to a level which would comply with appropriate management and risk practice [Croll, 2005]. Indeed poor spreadsheet modeling and human error in the design, logical computation, and control of spreadsheets can pose significant risks to a business. [Panko, 2000]

Undetected flaws in spreadsheets give unreliable management information which can lead to unsound business decisions and lost revenue. [Chadwick, 2003] Furthermore, uncontrolled and untested spreadsheets can leave an organization open to fraud and tampering, as well as difficulties in demonstrating regulatory compliance.

## 2. CONTEXT

Many organisations have seen the operational risks they face change as the requirements for speed and flexibility in product development or innovation are measured against an increasing requirement to be compliant, to operate in terms of standard or best practices, and to mitigate any exposures against potential risks. The pace of change and innovation in products requires organisations to act quickly. Systems solutions would be ideal although more and more end-user computing applications (EUCA) offer the simplest development solution to a computational or model based task.

At the same time end-user tools are often a cause for concern because, by their nature, they are sole end-user developed, fall outside of the standard systems development process for the organisation, are often not properly scoped, and are developed at varying levels of complexity, and supported by varying levels of documentation. Such spreadsheets, while they can gain quite a good deal of operational status and value in the activity of the organisation may often remain un-checked, un-validated and perhaps be unsafe. The risk matrix below illustrates how risk and probability are related over changing circumstances.

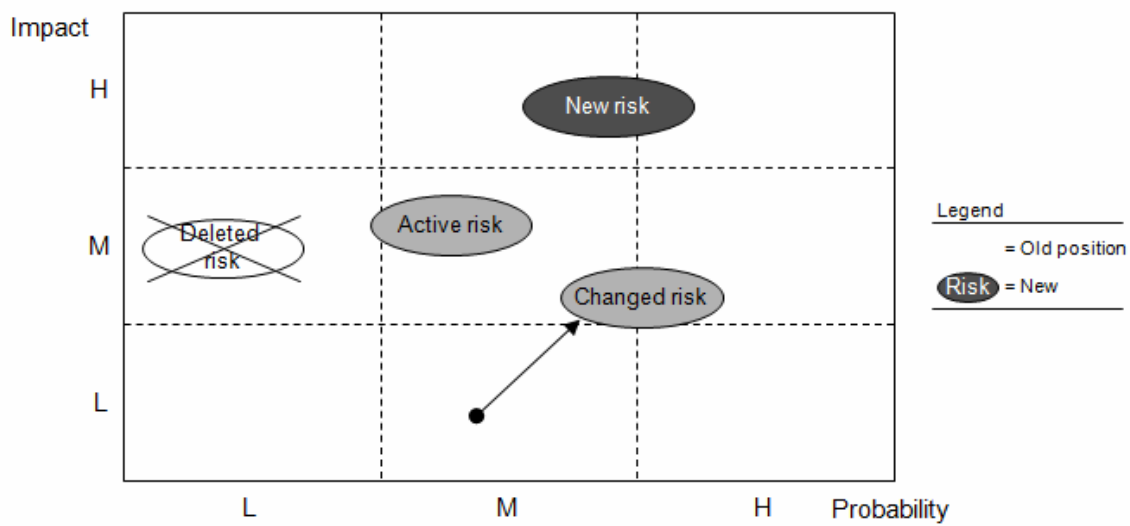

**Figure 1 Risk Matrix**

## 3. SPREADSHEET PARADIGM

In business today spreadsheets play a fundamental role in the management systems of organisations. Much of the key information used by businesses in summary decision-making, financial reporting, and project management is underpinned by countless thousands of spreadsheets generated by many individuals at all levels within organisations. Although we may imagine that formal systems solutions provide the infrastructure and reporting for systems processing and routine operational activity, more and more we come to recognize that the speed, and flexibility of the spreadsheet is uniquely attractive where different models are to be explored, different scenarios to be run, different reports to be compiled and summarized. Spreadsheets help managers and end-users develop professional looking reports and models to meet pressing client needs. That said untested and undocumented ad-hoc solutions may cause more problems down the line. Best practices or recommended or standard practices seem to be necessary [Nick Read and Jonathan Batson, 1999]; [Panko, 2006]; [Chadwick, 2007]; [O'Beirne, 2005]. A training and certification programme for managers and end-users, along with a whole series of internal awareness raising measures, as well some internal controls and policies may go some way to help to deal with these challenges.

## 4. PROGRAMME OVERVIEW

The programme goals from the Spreadsheet Safe™ Syllabus [Q-Validus, 2008] characterise the programme as follows:

> **Spreadsheet Safe™** is a certificate programme to help spreadsheet end-users and organisations manage and maintain safe spreadsheets. Candidates shall be able to setup, arrange and present their spreadsheets based on standard best practices in spreadsheet design, end-use and control. Spreadsheet Safe™ Candidates shall achieve well set out, error-free spreadsheets, which show a clear history and ownership path. Candidates shall also be able check their work for input accuracy, breaking down formulas into smaller more auditable parts, as well as making units of measure explicit for ease of update. Spreadsheet Safe™ Candidates shall routinely check their work for calculation and output accuracy, as well creating and validating formulas. Candidates shall also recognize common spreadsheet errors. Spreadsheet Safe™ Candidates shall be able to create charts, and choose the most suitable chart types in order to illustrate data accurately and to enhance meaning. Spreadsheet Safe™ Candidates are required to work in an informed and responsible way, and to appreciate the importance of spreadsheet review, audit and validation. Spreadsheet Safe™ Candidates shall also be aware that spreadsheets may be auditable as part of regulatory requirements. Candidates shall also have due regard to pertinent Health and Safety issues with regard to using computers. Spreadsheet Safe™ helps Candidates demonstrate their skill and awareness in working carefully and productively with spreadsheets.

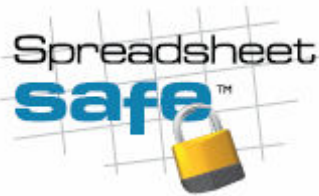

**Figure 2: Spreadsheet Safe Logo**

The Spreadsheet Safe™ certification programme has been developed by Q-Validus, in conjunction with its learning and assessment partners, BPP Learning Media and BTL Learning & Assessment. The Spreadsheet Safe™ programme provides:

- Spreadsheet Safe™ Training Manual
- One day interactive training session
- Spreadsheet Safe™ Online Training
- Spreadsheet Safe™ Certification Test
- Spreadsheet Safe™ Certificate

## 5. STANDARD PRACTICES

The term 'standard practice' in this paper is used to express sound, and reasoned spreadsheet work practices, which are themselves underpinned by informed attitudes and a reasonable knowledge about spreadsheet risks and productivity. The term is used in a distinctive way to other qualifiers like 'best', 'good' or 'recommended' practices. The reason for this is to suggest that our knowledge about spreadsheet risks and spreadsheet productivity is such that standard practice approaches may now be necessary since the risks seem accepted, plausible and evidenced, and while spreadsheet usage remains so common. In the paper it is suggested that standard practices should apply generally, given the inherent risks entailed, or productivity gains lost, where such practices are absent. A training and certification programme is one feature of this kind of work effort. In this regard, the verbs used to describe and express the skills, knowledge, and sensitivities of the competent, careful and productive spreadsheet end-user provide insights about how best to proceed to establish some evolving standard practices. A simple process in the paper was used as follows.

## 6. PROCESS STEPS

The first step was to review the kinds of cognitive verbs often used to describe what may be considered as standard spreadsheet practices, and then think about how the kinds of competencies or knowledge, and implicit behaviours, entailed by the verbs may be suitable to training and measurement. The reason verbs were chosen as distinct from nouns is that verbs are more important in the measurement skills or knowledge. A persistent group of verbs emerged in the various practice definitions reviewed, and these were examined and listed. The verbs were then categorised in terms of Bloom's Taxonomy of Hierarchical Levels, [Bloom, Krathwohl, 1956], categorised and illustrated over the hierarchical levels, and then arranged using an overlay scheme to qualify and distinguish the practices entailed by distinguishing the standard practice descriptive verbs in terms of knowledge, skill, behaviour and responsibility. The grid scheme considered:

- things which can be known
- skills which may be functional
- behaviours which can be learned or re-enforced
- roles entailed which seem responsible

In this way, qualities and characteristics are anticipated for end users, and for the organisation, premised on the standard practice verbs which emerge. As an additional exercise, some of the verbs were inverted to their negative form to give insights about the apparent risks and loss to productivity, where less care or diligence is given. Also a somewhat arbitrary, although probably reasonable, risk rating value [High, Moderate, or Low] was applied to some inverted verbs. Finally some measurement and testing items were considered, which originate from the Spreadsheet Safe™ certification test bank, as a way to draw some inferences about how adequately testing / certification items may measure against standard practices, as part of a wider process to help organisations and end-users use spreadsheets in a more informed, productive and standardized way.

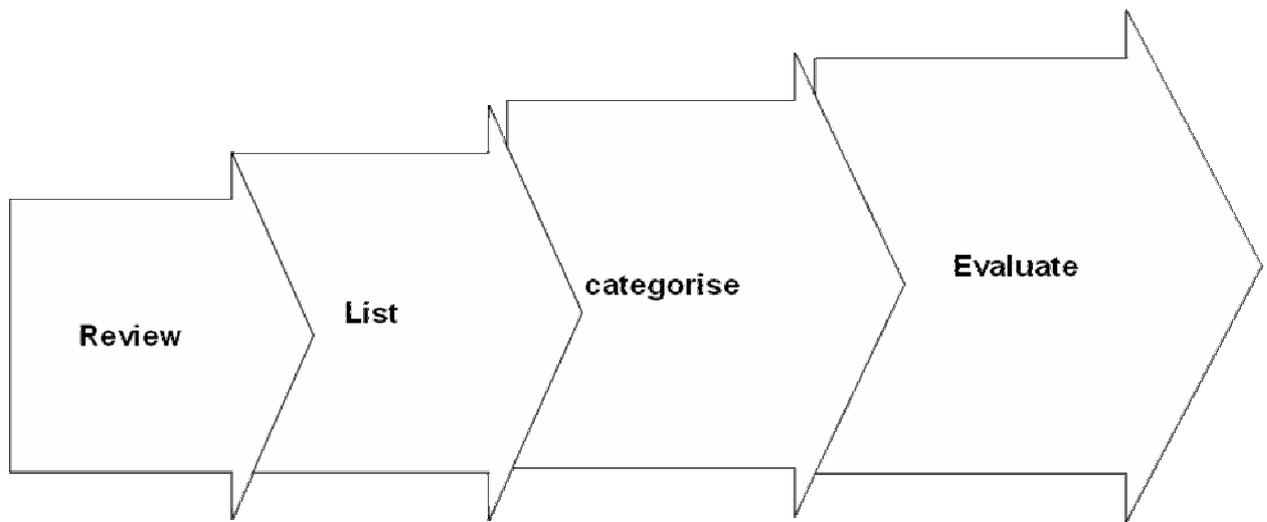

**Figure 3 Process Diagram**

The 'Review' process in the scheme above describes the first step in the evaluation process which was to review various standard practice descriptors and to begin to list the kinds of practice oriented verbs used.

## 7. STRONG COGNITIVE VERBS

The scheme below illustrates the strong cognitive verbs which emerge and persist from the literature review for spreadsheet practices. A frequency rating is not given or implied. Instead the table simply illustrates the kinds of standard practice verbs which feature consistently throughout the literature reviewed.

| know | understand | state | list | organise | save | verify | protect |
|---|---|---|---|---|---|---|---|
| determine | check | express | set out | lay out | backup | archive | recognise |
| simplify | use | precise | import | reconcile | examine | import | calculate |
| correct | cross-check | show | un-hide | separate | export | print | proof |
| control | maintain | switch | orient | submit | collaborate | remove | check |
| automate | manage | over-see | circulate | access | track | trace | recognise |
| sign-off | incorporate | interact | approve | notify | report | link | audit |
| think | design | implement | eliminate | modify | beware | arrange | build |
| audit | break-up | inventory | reference | trap | reflect on | highlight | distinguish |
| publish | log | archive | cooperate | own | define | document | format |
| control | validate | inspect | illustrate | test | reveal | replace | suppress |
| process | scope | specify | model | review | plan | list | clarify |
| avoid | calculate | conceptualise | refine | present | recognise | protect | document |
| communicate | determine | amend | separate | order | formalise | persist | colour |
| change | code | visualise | maintain | illustrate | review | refer | be aware |

**Table 1: Strong Cognitive Verbs**

## 8. BLOOM'S TAXONOMY OF HIERARCHICAL LEVELS

Bloom's Taxonomy of Hierarchical Levels [Bloom, Krathwohl, 1956], relates different cognitive levels for verbs in a cognitive value scheme as follows:

1. **Knowledge** [Where Candidates remember previously learned material.]
2. **Comprehension** [Where Candidates grasp meaning.]
3. **Application** [Where Candidates apply learning in a concrete way.]
4. **Analysis** [Where Candidates break down material so as to understand it.]
5. **Synthesis** [Where Candidates brings things together to make something new.]
6. **Evaluation** [Where Candidates weigh and judge value and application given a purpose.]

The standard practices verbs from Table 1 were reviewed and then included in the different categories to try and gain an estimate for the predominant cognitive levels.

| Knowledge | Know / Understand / lists / Define / Recognise / Document / Describe / Identify [**total 8**] |
|---|---|
| **Comprehension** | Determine / Think / Understand / State / Express / Show / Set out [**total 7**] |
| **Application** | Automate / Process / Change / Use / Maintain / Log / Calculate / Implement / Print / Import / Un-hide / Access / Notify / Save / Backup / Export / Reveal / Link / Replace / Refer / Protect / Build / Suppress / Colour / Reference / Re-use / Remove / Apply / Name / Create / Version / [**total 31**] |
| **Analysis** | Simplify / Avoid / Check / Cross-Check / Break-up / Precise / Switch / Analysis / Trap / Distinguish / Separate / Highlight / Isolate / Split out [**total 14**] |
| **Synthesis** | Communicate / Manage / Incorporate / Synthesis / Scope / Interact / Archive / Specify / Conceptualise / Co-operate / Refine / Organise / Lay out / Present / Order / Illustrate / Collaborate / Report / Formalise / Clarify / [**total 20**] |
| **Evaluation** | Correct / Control / Evaluation / Audit / Publish / Validate / Inspect / Examine / Orient / Approve / Eliminate / Reconcile / Submit / Test / Review / Track / Beware / Reflect on / Verify / Validate / Trace / Persist / Proof / Check / Examine / Comply / Own [**total 27**] |

**Table 2 Hierarchical Levels**

The scheme illustrates a large number of verbs (57%) categorised at the levels of **Analysis**, **Synthesis** and **Evaluation** in terms of Bloom's Taxonomy. These levels are characterised as higher order cognitive levels.

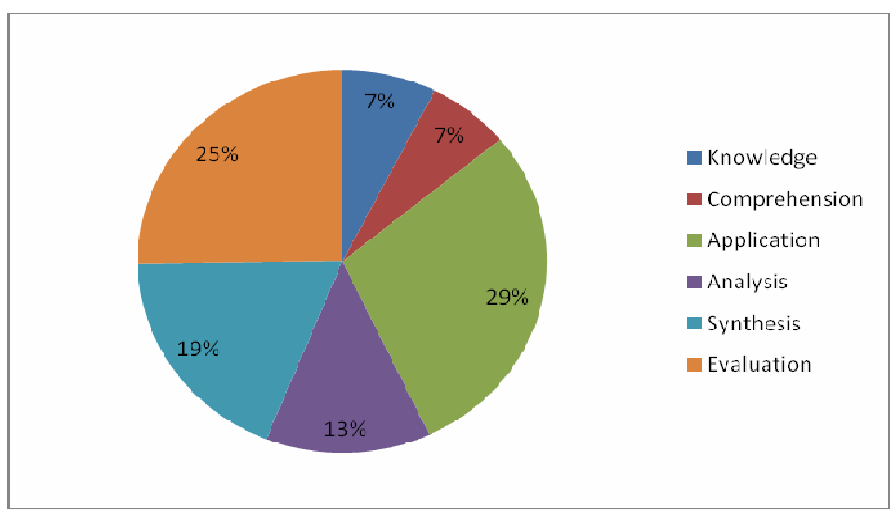

**Figure 4: Cognitive Levels Breakdown**

## 9. OVERLAY SCHEME

The verbs were then overlaid on a simple scheme to qualify and distinguish what was anticipated in the verbs as follows:

- KNOWLEDGEABLE - things which can be known
- FUNCTIONAL - skills which may be functional
- BEHAVIORAL - behaviours which can be learned / re-enforced
- RESPONSIBLE - roles predicated by the verbs

Qualities for functional skills, knowledge, behavioural or responsible outcomes entailed by the cognitive verbs were chosen, and then illustrated in a grid scheme. Some verbs illustrated characteristics which were functional but which also had a predominant bias towards the responsible or behavioural quadrants. A best estimate approach was applied as to where a verb best resided. The grid scheme qualified and distinguished the different standard practice verbs and associated them with their predominant qualities.

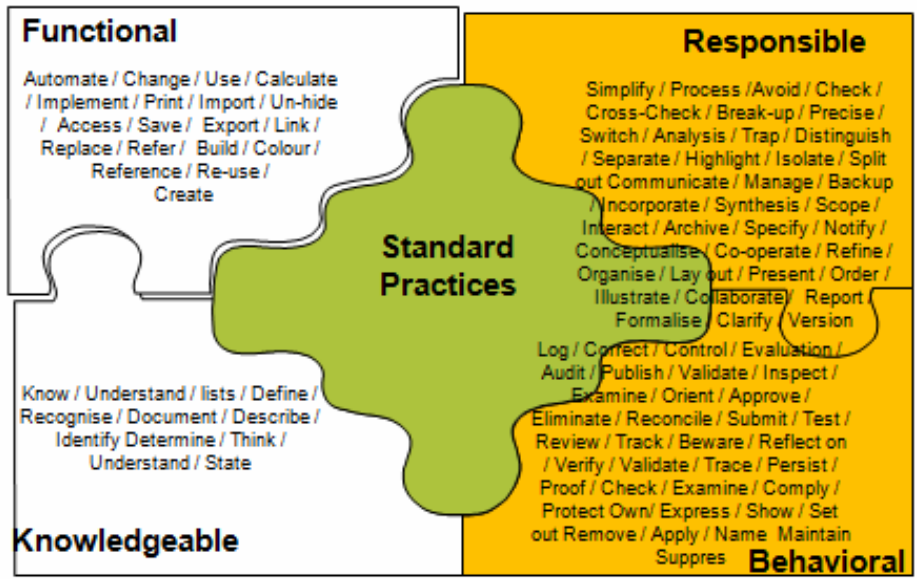

**Figure 5: Overlay Scheme**

It is a simple exercise and illustration. Although it can be seen that for the standard practices anticipated, quite a degree of the practice verbs are in the Responsible and Behavioural areas, more than the straight functional or knowledge areas. As such we can deduce the significance of training and measurement to promote attitudes and awareness around considered, careful and productive use over rote knowledge or routine application skills.

## 10. STRONG COGNITIVE VERB INVERSION EXTRACT

Some of the verbs were then inverted in a separate extract model to gain a sense of the impact when standard practices are not applied.

| unknown | misunderstand | unexpressed | non-compliant | disorganised | unsaved | imprecise |
| --- | --- | --- | --- | --- | --- | --- |
| undetermined | unrevised | uncontrolled | unclear | untraceable | unsaved | not archived |
| complicated | unmanaged | imprecise | vague | unreconciled | incorrect | invalid |
| un-scoped | unchecked | hidden | unprotected | unspecified | misleading | inaccessible |

**Table 3: Verb Inversion Extract**

Presented in the negative, the impacts seem startling. By considering spreadsheet applications competence distinctly from assumed spreadsheet practice standards, the casual inference is made that work practices may be at a high level for everyone in the organisation. All the while standard spreadsheet practices may not be in place, and even the default spreadsheet skills assumption may be erroneous.

## 11. HML RISK GRID

A risk rating scale was also applied to see how applicable the cognitive values anticipated by spreadsheet end-users reflect on the inverted verb scheme. A High / Moderate / Low risk assessment scheme is applied as follows:

| unknown | **M** | misunderstood | non-compliant | **H** | unprocessed | **M** | unstated | **M** |
| --- | --- | --- | --- | --- | --- | --- | --- | --- |
| undetermined | **M** | invalid | unclear | **M** | unclear | **M** | untracked | **H** |
| complicated | **M** | unmanaged | vague | **M** | imprecise | **H** | unaudited | **H** |
| incorrect | **H** | unchecked | un-hidden | **M** | hidden | **H** | un-shown | **H** |
| erroneous | **H** | misleading | unmaintained | **L** | untracked | **H** | ill-defined | **M** |

**Table 4: HML Grid [High / Moderate / Low outcome and impact risk probability]**

The risk rating scale is applied somewhat arbitrarily, although I was guided by criteria such as the stakes involved, the potential business impacts, and perceived productivity losses, which might be likely or possible when standard practices were not be applied.

## 12. MEASUREMENT TREATMENT

The verb inversion exercise and the risk rating scheme seem to show, however simply, that risk increases and productivity decreases when standard practice approaches associated with spreadsheets usage among end-users and organisations are ignored.

A valid measure to deal with this may be training and certification. In this way a Spreadsheet Safe™ Syllabus extract is used to consider the testing treatment, for some of the cognitive verbs already seen.

First a judgement is made about whether a knowledge-based (KB) or simulations-based (SIM) question item treatment approach is predominantly suitable. Then a short review of some questions items is given to see how adequately the items measure and reinforce some of the standard practices reflected in the Syllabus.

| CATEGORY | SKILL AREA | REF. | TASK ITEM | SIM | KB |
|---|---|---|---|---|---|
| **1.1 SETUP** | *1.3.1 Organize* | 1.1.2.4 | Apply strong password protection to spreadsheets, using mixed case, and non-alphanumeric characters (at least 8) to protect against unauthorised access. | | 1 |
| **1.2 INPUTS** | *1.2.1 Controls* | 1.2.1.1 | Place a single instance of a given constant (e.g.: conversion or tax rates) in separate cells. | 1 | |
| | | 1.2.1.2 | Break down more complex formulas into smaller component parts to help with readability, comprehension, and for ease of update. | 1 | 1 |
| **1.3 CALCULATE** | *1.3.1 Formulas* | 1.3.1.1 | Use mathematical and logical formulas and functions in a spreadsheet. | 1 | |
| **1.4 OUTPUTS** | *1.4.2 Charts* | 1.4.2.3 | Change the orientation of a chart so that all data series are visible. | 1 | |

**Table 5: Spreadsheet Safe Syllabus Extract**

## 13. ILLUSTRATIVE QUESTION ITEMS AND CONGRUENCE

The following question item from the Spreadsheet Safe™ test bank is based on knowledge about password protection. An automated questions item with a knowledge-based approach, seems to measure against the cognitive level anticipated here reasonably well. A simulations-based approach could equally be used.

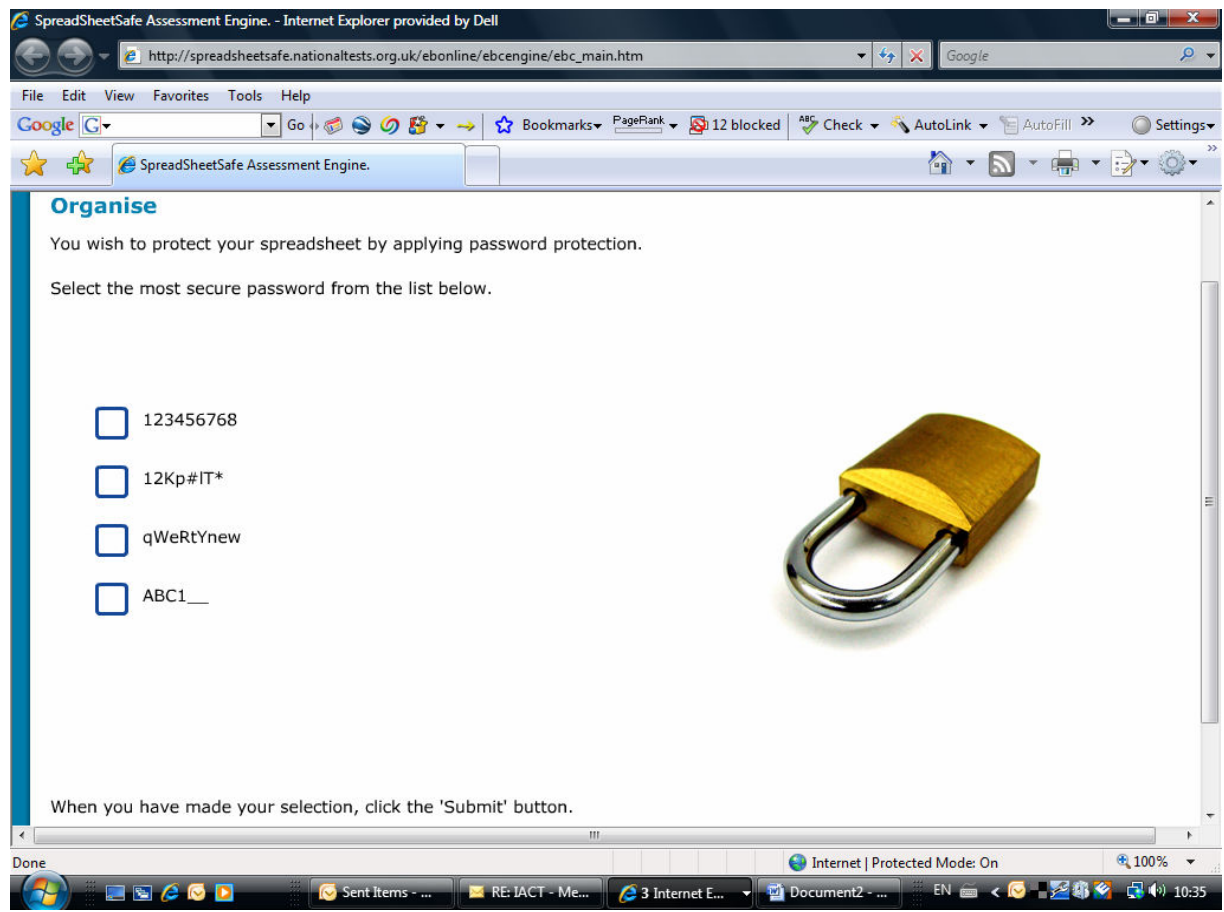

**Figure 6 : Question Item**

In the password protection item above we can see how strong password protection can be promoted and applied in assessment. As such, we can judge this item to be congruent with the measuring point expressed in the Syllabus.

The following question item from the Spreadsheet Safe™ test bank is based on the Candidate's knowledge or ability to detect or recognise a named range, and appreciate how it might be used.

**Figure 7: Question Item**

In the above item we can see how a named range can be distinguished as a separate constituent element in the formula. As such we can judge this item to be congruent with the measuring point expressed in the Syllabus and help promote standard practice.

The following question item from the Spreadsheet Safe™ test bank measures the Candidate's ability to correct precedent cells.

**Errors**

Correct the formula in the formula bar below to include the correct precedent cells for the Country Break tour from 2004 to 2007

| | G10 | | fx =SUM(B10:E10) | | | | | |
|---|---|---|---|---|---|---|---|---|
| | A | B | C | D | E | F | G | H |
| 1 | Suntime Tours Ltd | | | | | | | |
| 2 | | Current and previous years | | | | | | |
| 3 | | | | | | | | |
| 4 | Tour Code | Tour Description | 2004 | 2005 | 2006 | 2007 | Total | |
| 5 | ABL | Anywhere but London | 89350 | 59201 | 43260 | 70354 | 191,811 | |
| 6 | CIT1 | City Break short | 89372 | 73507 | 66525 | 94476 | 229,404 | |
| 7 | CIT2 | City Break Long | 23163 | 83566 | 44758 | 42049 | 151,487 | |
| 8 | CUL1 | Culture & Museums | 48994 | 96325 | 27707 | 16079 | 173,026 | |
| 9 | DEC1 | Decadent Luxury | 21243 | 32726 | 11498 | 60736 | 65,467 | |
| 10 | DEC2 | Country Break | 98624 | 71734 | 60388 | 20219 | 230,746 | |
| 11 | | | | | | | | |
| 12 | | | | | | | | |

Reset formula

**Figure 8: Question Item**

In the above item we can see how the correct precedent cells item is measured with a practical simulations-based testing item. Correcting what is incorrect may help to promote standard practice.

The following question item from the Spreadsheet Safe™ test bank measures the Candidates awareness of the need to re-orientate a chart in order to meaningfully communicate in a report to colleagues, or to an audience.

**Figure 9: Question Item**

The orientation of the chart changes so that the information in the chart is clearly communicated. This promotes a standard practice approach and develops Candidate sensitivity to chart selection.

## 14. SUMMARY AND CONCLUSIONS

When we consider the knowledge, skills and behaviours anticipated in the descriptive verbs which are used with standard practices in spreadsheets for end-users we can see a significant reliance on responsible behaviours, conscientious attention to detail, as well as the assumed applications skills of working out calculations reliably and routinely. Often the awareness and sensitivity required to operate at a standard practices level seem predominantly to be higher order cognitive level knowledge and skills such as those contained within the Analysis, Synthesis, and Evaluation levels in Bloom's Taxonomy. When we consider that business critical spreadsheets and productivity are at stake for organisations, and when we invert some standard practice illustrative verbs, and then rate and measure the risks and productivity costs that are themselves inherently tied up in spreadsheets and how they are used, we can appreciate that training and certification, perhaps along with some other awareness raising measures and organisational controls, may help to promote standard spreadsheet practices and approaches.